\documentclass[lettersize,journal]{IEEEtran}
\usepackage{amsmath,amssymb,amsfonts}
\usepackage{algorithmic}
\usepackage{array}
\usepackage[caption=false,font=normalsize,labelfont=sf,textfont=sf]{subfig}
\usepackage{textcomp}
\usepackage{stfloats}
\usepackage{url}
\usepackage{color}
\usepackage{verbatim}
\usepackage{graphicx}
\usepackage{cite}
\newtheorem{remark}{Remark}
\newtheorem{assumption}{Assumption}
\newtheorem{lemma}{Lemma}
\newtheorem{definition}{Definition}

\newtheorem{theorem}{Theorem}

\def\BibTeX{{\rm B\kern-.05em{\sc i\kern-.025em b}\kern-.08em
    T\kern-.1667em\lower.7ex\hbox{E}\kern-.125emX}}
\usepackage{balance}
\begin{document}
\title{Adaptive Control with  Guaranteed Transient Behavior and Zero Steady-State Error  
	for Systems with Time-Varying Parameters}
 \author{Hefu Ye, and Yongduan Song, \IEEEmembership{Fellow, IEEE}
 	\thanks{This work was supported by the National Natural Science Foundation		of China under grant (No.61991400, No.61991403, No.61860206008, and No.61933012). (Corresponding Author: Yongduan Song.)}		\thanks{H. F. Ye   is with Chongqing Key Laboratory of Autonomous Systems, Institute of Artificial Intelligence, School of Automation, Chongqing University, Chongqing 400044, China, and also with Star Institute of Intelligent Systems (SIIS), Chongqing 400044, China.   (e-mail:    yehefu@cqu.edu.cn).} 	\thanks{Y. D. Song is with Chongqing Key Laboratory of Autonomous Systems, Institute of Artificial Intelligence, School of Automation, Chongqing University, Chongqing 400044, China. 		(e-mail:   		ydsong@cqu.edu.cn).} }

\maketitle

\begin{abstract}
	It is nontrivial to achieve global zero-error regulation for uncertain nonlinear systems. The underlying problem becomes even more challenging if mismatched uncertainties and unknown time-varying control gain are involved, yet certain performance specifications are also pursued. In this work, we present an adaptive control method, which, without the persistent excitation (PE) condition, is able to ensure global zero-error regulation  with guaranteed output performance for parametric strict-feedback systems involving fast time-varying parameters in the feedback path and input path.  The development of our control scheme benefits from     generalized $t$-dependent and  $x$-dependent functions, a novel coordinate transformation and  ``congelation of variables" method. Both  theoretical analysis and numerical simulation verify the effectiveness and benefits of the proposed method.
\end{abstract}


\section{Introduction}
\label{sec:introduction}
\IEEEPARstart{W}{e}  consider the following SISO nonlinear systems with fast time-varying parameters\cite{chenkaiwen}
\begin{equation}\label{system}
\left\{\begin{array}{ll}
\dot{x}_1=\phi_1^{\top}(  {x}_1)\theta(t)+x_{2}\\ 
~~~\vdots\\
\dot{x}_i=\phi_i^{\top}(\underline {x}_i)\theta(t)+x_{i+1}\\ 
~~~\vdots\\
\dot{x}_n=\phi_n^{\top}(\underline{x}_n)\theta(t)+b(t)u\\
y=x_1
\end{array}\right.
\end{equation}
where $\underline{x}_i=[x_1,\cdots,x_i]^{\top}\in \mathbb{R}^i$ is the state vector, $u\in\mathbb{R}$ is the input, $y\in\mathbb{R}$ is the output. The regressors $\phi_i:\mathbb{R}^i\rightarrow\mathbb{R}^q,~i=1,\cdots,n,$ are smooth mappings and satisfy $\phi_i(0)=0.$ $\theta(t)\in\mathbb{R}^q$ and $b(t)\in\mathbb{R}$ satisfy the following Assumptions\cite{chenkaiwen}.
\begin{assumption}[Bounded parameters]
	The parameter $\theta(t)$ is piecewise continuous and $\theta(t)\in \Theta_0,$ for all $t\geq 0$, where $\Theta_0$ is a compact set. The ``radius" of $\Theta_0$ is assumed to be known, while $\Theta_0$ can be unknown.
\end{assumption}

\begin{assumption} [Sign-definite parameter]
	The control gain $b(t)$ is bounded away from zero in the sense that there exists a constant $\ell_b$, such that $\text{sgn}(\ell_b)=\text{sgn}(b(t))\neq 0$ and $0<|\ell_b|\leq |b(t)|$, for all $t\geq 0$. The sign of $b(t)$ is known and does not change.
\end{assumption}

Stabilization of system (\ref{system}) satisfying  Assumptions 1-2 is originally investigated in \cite{chenkaiwen,chenkaiwenIFAC,chenkaiwenACC}, where it is shown that asymptotic stability can be achieved by the so-called \textit{congelation of variables} method and both full state feedback and partial state feedback approaches are considered. By ``congelation of variables" it means that the time-varying $\theta(t)$ can be substituted  by constant $\ell_{\theta}$ ($\ell_{\theta}$ can be regarded as the average of $\theta(t)$) to avoid unnecessary time derivatives while not destroying the certainty equivalence principle \cite{krstic}. It is noted that if the parameter $\theta(t)$ in (\ref{system}) is unknown but constant, numerous adaptive control results have been reported in literature during the past decades, including the well-known adaptive backstepping control, robust and adaptive control, adaptive observers, immersion and invariance adaptive control, neural adaptive control, etc. (see  \cite{krstic,WenChangyun,Astolfi,Ioannou-PE,hill,CAA-wanghuanqing,CAA-tongshaocheng} and the references therein). 

However, real-word engineering systems with fast time-varying parameters are frequently encountered. For instance, the value of a circuit resistor might change with temperature, some morphing aerial vehicles are normally designed with varying structures and parameters in order to complete some specific tasks, where the parameters might change with time or system states swiftly \cite{Quadrotor,nature}. For this type of systems, traditional adaptive methods might not be able to ensure desired control performance in terms of transient behavior and convergence accuracy, or even unable to maintain system stability. Efforts have been made (see, for instance \cite{Goodwin-PE} and  \cite{Kreisselmeier-PE}) in developing adaptive control methods with the aid of the persistence of excitation (PE) to achieve exponential stability of  linear time-varying systems. In \cite{Goodwin-nonPE}, it is  shown that the PE condition is not necessary to stabilize a linear time-varying system. The results in \cite{song1992} and \cite{Songa} implement  the asymptotic/exponential tracking of robotic systems with/without time-varying parameters.
In \cite{kokotovic,Marino,Marino-TAC}, along with observer based adaptive control,  projection algorithm is proposed to ensure the boundedness of slow time-varying parameter estimate. In the context of adaptive control for time-varying nonlinear systems, the work  \cite{Huang2018} explorers a soften sign function based approach to deal with unknown time-varying parameters. Recently, an elegant method based on  ``congelation of variables" is proposed in \cite{chenkaiwen,chenkaiwenIFAC,chenkaiwenACC}  to asymptotically stabilize a class of nonlinear system with fast time-varying parameters, which is further extended to address multi-agent systems in \cite{ChenYY} and \cite{ChenYY2}. Thus far,  meaningful results on adaptive control of  systems with unknown and fast time-varying parameters are still limited, rendering the underlying problem interesting yet challenging.

In this note, we address the stabilization problem of fast time-varying system as described in (\ref{system}) and our goal is to achieve zero-error full state regulation and at the same time maintaining global  output performance, \textit{i.e.,} regulating each state to zero asymptotically and meanwhile confining the convergence process of the output within an prescribed boundary. Our development consists of three major steps:  $i)$ disassociating the recursive controller design
from the initial condition of system (\ref{system}) via two generalized functions and a novel coordinate transformation; $ii)$ designing adaptive laws to estimate fast time-varying parameters involved in the constrained systems; and $iii)$ separating the lumped nonlinear terms and exploiting additional nonlinear damping terms in each virtual control input to finally offset the undesired perturbations caused by unknown time-varying control gain. With this comprehensive treatment, output convergence transient behavior is well preset and asymptotic (zero-error) regulation is achieved in the presence of mismatched time-varying uncertainties.

Unlike most prescribed performance control methods that only achieve uniformly ultimately bounded (UUB) for nonlinear systems with unknown but constant parameters  \cite{Wang2017,Wang2010,Benchlioulis,Zhao-stateMIMO,Zhao-stateAuto,zhang,zhangcyber-global,CAA-sunchangyin,CAA-tongshaocheng2,huangxiuchai,zhoushuyan,caoye}, the proposed method ensures zero-error stabilization and global output performance for systems with fast time-varying parameters and mismatched uncertainties.

\section{preliminaries}\label{sectionII}
\subsection{Two Useful functions \& Coordinate transformation} \label{2.1}
Before presenting the control algorithm, we introduce two useful functions and a novel coordinate transformation, which plays important roles in control design.
	\begin{definition}
The generalized performance function $\beta(t)$ satisfies the following properties: 
\begin{itemize}
	\item $\beta(t):[0,\infty)\rightarrow \mathbb{R}^{+}$ is a $n$-times  differentiable function;
	\item $\beta(0)=1$ and $\lim_{t\rightarrow+\infty}\beta(t)<1$;
	\item $\beta(t)\in\mathcal{L}_{\infty}$ and $ \dot\beta(t)\in\mathcal{L}_{\infty},~\forall t\in[0,+\infty)$.
\end{itemize}
	\end{definition}
\begin{remark}\label{remark-beta}
	There are many (in fact, infinite number of) functions
	that satisfy the aforementioned properties. For
	example,  
	\begin{equation}\label{beta}
	\begin{array}{l}
	\beta(t)=\left\{\begin{array}{l} (1-\beta_{\infty})\left(\frac{T-t}{T}\right)^n+\beta_{\infty},~0\leq t<T\\
	\beta_{\infty},~~~~~~~~~~~~~~~~~~~~~~~ ~~t\geq T;
	\end{array}\right. 
	\end{array}
	\end{equation}
	where $\beta_{\infty}=\lim_{t\rightarrow+\infty}\beta(t)$, $T>0$ is a constant and $n$ is the system order.  Note that the performance function is not necessarily monotonically decreasing, which might 	be advantageous in various applications, \textit{e.g.,} when the system time-varying parameter changes strongly or the system is perturbed by some calibration so that a large error would enforce a large input action.
\end{remark}
\begin{definition}
		The generalized normalized function $\psi(x)$ satisfies the following properties: 
\begin{itemize}
	\item $\psi(x):\mathbb{R}\rightarrow (-1,1)$ is a monotonically increasing  and $n$-times  differentiable function;
	\item $\lim_{x\rightarrow\pm\infty}\psi(x)=\pm1$ and $ \psi(0)=0$;
	\item $\psi'(x)$ is bounded below by a positive constant over $[0,\infty)$, where $\psi'(x)=\frac{d\psi}{dx}$.
\end{itemize}
	\end{definition}
\begin{remark}\label{remark-psi}
	We list two choices for $\psi(x)$ as follows:
	\begin{equation}\label{psi}
	\begin{array}{l}
	\psi(x)=\frac{x}{\sqrt{x^2+1}}; ~~
	\psi(x)=\tanh(x),
	\end{array}
	\end{equation}
and for the above two choices, we have:
 	\begin{equation}\label{dpsi}
 \begin{array}{l}
 \psi'(x)=\frac{1}{(\sqrt{x^2+1})^{3/2}}; ~~
 \psi'(x)=\text{sech}^2(x).
 \end{array}
 \end{equation}
 Denoting the inverse function by $\psi^{-1}$, it is seen that\footnote{Property 1 and Property 2 of $\psi(x)$ ensue that $\psi_x$ is positive and invertible for all $x \in \mathbb{R}$.} 
 \begin{equation}\label{property-psi}
 	\begin{aligned}
 	&\psi'(x)>0,  ~~~~\psi_x\triangleq\frac{\psi}{x}>0, \\
 	&\psi^{-1}(\beta(0))=\psi^{-1}(1)=+\infty.
\end{aligned}
 	\end{equation} 
\end{remark}  
 Making use of such $\beta(t)$ and $\psi(x)$, we construct the following   coordinate transformation function to  enable the properties on $z $ and $x $ as stated in  Lemma \ref{lemma1}.
\begin{equation}\label{z1}
z (\beta,\psi )=\frac{\beta(t) \psi(x )}{\beta^2(t) -\psi^2(x )}.
\end{equation}  
\begin{lemma}\label{lemma1}
	For any $\beta(t)$ as defined in Section-\ref{sectionII}  and $z $ as defined in  (\ref{z1}), if $\forall t\geq 0$, $z \in \mathcal{L}_{\infty}$, then it holds that $-\psi^{-1}(\beta)<x <\psi^{-1}(\beta)$. 
\end{lemma}
\textit{Proof:} We first consider the moment when $t=0$. According to  $\beta(0)=1$ and   $\psi(x)\in(-1,1)$, we know that $\beta(0)-\psi(|x(0)|)>0$, \textit{i.e.}, $|x(0)|<\psi^{-1}(\beta(0))$. 
Next, we continue the proof by contradiction. Note that $z\in\mathcal{L}_{\infty}$ implies $\beta(t)-\psi(x)\neq 0$.  Assume that $\exists~ t\in(0,\infty)$ such that $|x(t) |\geq\psi^{-1}(\beta(t) )$, \textit{i.e.}, $\beta(t)-\psi(|x(t)|)\leq0$. As a result, by recalling that $\beta(0)-\psi(x(0))>0$, we have $\exists ~t_1\in(0,t]$ causes $\psi(|x (t_1)|)=\beta(t_1)$, and therefore yields an unbounded $z_1$, which, however,  contradicts the premise $z \in\mathcal{L}_{\infty}$. This completes the proof. $\hfill\blacksquare$ 

	This coordinate transformation introduced in (\ref{z1}) appears as a more straightforward approach compared to the tuning function modified transformation \cite{zhangcyber-global} and the multiple cascade transformation\cite{zhaoTAC2021}, by reason of its simple structure, smoothness and nonsingularity.  

\subsection{Control Objective}
 The control objective is to design an adaptive control law such that the closed-loop system is asymptotically stable, while the system output is always confined within a prescribed performance funnel $\digamma_{\beta(t)}$. Furthermore, the boundary of $\digamma_{\beta(t)}$ is $\beta(t)$, which can be pre-defined at user's will,  irrespective of initial conditions.
\begin{remark}
If we choose a function $\beta(t)$ with an exponential decay rate, \textit{e.g.,} $\beta(t)=(1-\beta_\infty)e^{-t}+\beta_{\infty}$.
By qualitative analysis,  $\psi^{-1}(\beta)$ is a function that increases monotonically as $\beta\rightarrow \infty$, and $\beta(t)$ is a function that  decays exponentially as  $t\rightarrow\infty$, thus $\psi^{-1}(\beta)$ is a function that decays exponentially as $t\rightarrow\infty$ and $\psi^{-1}(\beta(0))\rightarrow\infty$.  Therefore,  $|x |<\psi^{-1}(\beta)$ implies that there exist some positive constants $l_1$, $l_2$ and $\epsilon$ such that  $|x(t)|<l_1e^{-l_2t}+\epsilon$ for any $x(0)$, 
resulting in that the system output converges at least $e^{-l_2t}$ exponentially fast to the corresponding set. Similarly, if we choose $\beta(t)$ as defined in (\ref{beta}), one can find that the system output converges to a prescribed set at a prescribed time $T$, a favorable feature in practice.
\end{remark}

\section{Motivating Example}\label{section3}
Consider the following first-order system\footnote{For simplicity, arguments of functions are sometimes omitted if no confusion is occur.}
\begin{equation}\label{scalar-system}
\dot{x} =b(t)u+\theta(t)x 
\end{equation} 
where $x $ is the state, $u$ is the control input, $\theta(t)\in\mathbb{R}$ satisfies Assumption 1, and $b(t)\in\mathbb{R}$ satisfies Assumption 2. 

 By using the coordinate transformation (\ref{z1}), we can convert (\ref{scalar-system}) into the following $z$-dynamics 
\begin{equation}\label{dz1-example}
\dot z =\Pi (x,t) \dot{x} + \Psi (x ,t) 
\end{equation}
with
\begin{equation*} 
\small{\begin{aligned}
	&\Pi(x,t) =  \frac{(\beta^2(t)-\psi^2(x ))^2\beta(t)\psi'(x )+2\psi^2(x )\psi'(x )\beta(t)}{(\beta^2(t)-\psi^2(x ))^2}\\ 
	& \Psi(x,t) =  \frac{\dot{\beta}(t)\psi(x )(\beta^2(t)-\psi^2(x ))-2\beta^2(t)\dot\beta(t) \psi(x )}{(\beta^2(t)-\psi^2(x ))^2},
	\end{aligned}}
\end{equation*} 
where $\Psi $ and $\Pi $ are  known time-varying smooth functions and   are bounded as long as $z $ is bounded. In addition, $\Pi >0$ for  $\forall z \in\mathcal{L}_{\infty}$.  These facts  ensure the controllability  of (\ref{dz1-example}).
 Motivated by \cite{chenkaiwen}, we design  $u=\hat\rho\bar u$, with $\hat\rho$ being an ``estimate" of $\frac{1}{\ell_b}$, and $\bar u$ being the compensating signal to be specified later, then (\ref{dz1-example}) can be written as
\begin{equation}\label{dz1-example-1}
\begin{aligned}
\dot z 
=&\Pi\left(\bar{u}+\hat\theta x +(\theta(t) -\ell_\theta)x +(b(t)-\ell_b) \hat\rho\bar u\right.\\
&+\left.(\ell_\theta-\hat\theta)x -\ell_b\left(\frac{1}{\ell_b}-\hat\rho\right)\bar{u}+\frac{\Psi}{\Pi} \right)
\end{aligned}
\end{equation}
where $\hat\theta$ is an ``estimate" of $\ell_\theta$, $ {\Psi}/{\Pi}\in\mathcal{L}_{\infty}$ for $\forall z\in\mathcal{L}_{\infty}$. Note that $\ell_\theta$ and $\ell_b$ are unknown constants, which can be regarded as the ``average" of $\theta(t)$ and $b(t)$, respectively.  Consider  the Lyapunov function candidate 
\begin{equation}\label{V}
V=\frac{1}{2}z ^2+\frac{1}{2\gamma_{\theta}}(\ell_\theta-\hat\theta)^2+\frac{|\ell_b|}{2\gamma_{\rho}}\left(\frac{1}{\ell_b}-\hat\rho\right)^2.
\end{equation}
Then,    the  derivative of (\ref{V}) along the trajectory of (\ref{scalar-system}) becomes
\begin{equation}\label{dV}
\begin{aligned}
\dot V=&\Pi\left(z\bar u+z\hat\theta x+z\Delta_{\theta}x+\frac{\Psi}{\Pi}z\right)\\
&+\Pi z\Delta_{b}\hat\rho\bar u+ \frac{1}{\gamma_{\theta}}(\ell_\theta-\hat\theta)(\gamma_{\theta}z\Pi x-\dot{\hat{\theta}})\\
&-\frac{|\ell_b|}{ \gamma_{\rho}}\left(\frac{1}{\ell_b}-\hat\rho\right)\left(\gamma_{\rho} \text{sgn}(\ell_b)z\Pi \bar{u}+\dot{\hat{\rho}}\right)
\end{aligned}
\end{equation}
where $\Delta_{\theta}=\theta(t)-\ell_\theta$ and $\Delta_{b}=b(t)-\ell_b.$ 
The  last two lines of (\ref{dV}) will be canceled by the following adaptive laws:
\begin{equation}\label{d_hat_Theta}
\dot{\hat\theta}(x,\beta)=\gamma_\theta z \Pi x,
\end{equation}
\begin{equation}\label{d_hat_rho}
\dot{\hat\rho}=-\gamma_\rho\text{sgn}(\ell_b) z \Pi\bar{u}.
\end{equation}
\begin{remark}\label{remark1}
	Note that $z (x ,\beta)$ as defined in (\ref{z1}) is a   smooth function and $x =0\Leftrightarrow z =0$, thus we can directly express $x  $ as $x  =W (x ,\beta)z $ by using Hadamard's Lemma (see \cite{chenkaiwen,chenkaiwenACC,chenkaiwenIFAC,hadamard}), where $W (x ,\beta)$ is a bounded smooth mapping for every bounded $z $. As a matter of fact, here $W =\frac{x  }{z }=\frac{x(\beta^2-\Psi^2) }{\beta\Psi }=\frac{ \beta^2-\Psi^2  }{\beta\Psi_x }\in\mathbb{R}^{+}$.
\end{remark} 

 According to Remark 1 and formula (\ref{property-psi}),  the perturbation terms in the first line of (\ref{dV}) can be rewritten as  
\begin{equation}
\begin{aligned}
z\hat\theta x+\frac{\Psi}{\Pi}  z  =&z\hat\theta x+\frac{\Psi_x}{\Pi}  zx= \left(\hat\theta+\frac{\Psi_x}{\Pi} \right) W(x,\beta)z^2\\
z\Delta_\theta x=&\Delta_\theta W(x,\beta)z^2.
\end{aligned}
\end{equation}    
By applying Young's inequality, then
\begin{equation}
\begin{aligned}
\left(\hat\theta+\frac{\Psi_x}{\Pi} \right) W(x,\beta)z^2  
\leq  \frac{1}{2} \left(\hat\theta+\frac{\Psi_x}{\Pi} \right)^2 W^2z^2
+\frac{1}{{2} } z^2,
\end{aligned}
\end{equation}
\begin{equation}\label{dz2.1}
\begin{aligned}
\Delta_\theta W (x,\beta)z^2 
\leq  \frac{1}{2} \delta_{\Delta_{\theta}}W^2z^2
+\frac{\delta_{\Delta_{\theta}}}{{2} }z^2,
\end{aligned}
\end{equation}
where $\delta_{\Delta_\theta}\geq |\Delta_\theta|$ is the  ``radius" of the compact set of $\theta(t)$.
Now consider $\bar u$ with a nonpositive nonlinear gain as
\begin{equation}\label{u1}
\begin{aligned}
\bar u=&- \left(\frac{k}{\Pi}+\frac{1}{2}\left( {\delta_{\Delta_\theta}} + {1} \right)+\frac{W^2}{2}\left(\hat\theta+\frac{\Psi_x}{\Pi} \right)^2\right)z\\
=&-\kappa(x,\beta,\hat\theta) z
\end{aligned}
\end{equation}
where $k>0$.

 We are now in the position to state the following theorem.
\begin{theorem}\label{theorem0}
	System (\ref{scalar-system}) with the control law (\ref{u1}) and the parameter update laws (\ref{d_hat_Theta}) and (\ref{d_hat_rho}) is globally asymptotically stable. Furthermore,  the state $x(t)$ is always confined within the prescribed performance funnel $\digamma_{\beta}:=\left\{(t, x)\in \mathbb{R}_{\geq0}\times \mathbb{R}\big| |{x(t)}|/\psi^{-1}(\beta(t))<1\right\}$, and ultimately converges to zero. Furthermore,  $\lim_{t\rightarrow\infty}\hat{\theta}$ and $\lim_{t\rightarrow\infty}\hat{\rho}$  exist  (although not necessarily equal to $\ell_\theta$ and $1/\ell_b$, respectively). In addition, the control input and update laws remain uniformly bounded over $[0,\infty)$. 
\end{theorem}
\textit{Proof:}  Substituting  (\ref{u1}) into (\ref{dV}),   yields 
\begin{equation}\label{dV-V}
\begin{aligned}
\dot V\leq& -kz^2+\Pi\left( z\Delta_\theta x-\frac{1}{2} \delta_{\Delta_{\theta}}W^2z^2
-\frac{\delta_{\Delta_{\theta}}}{{2} }z^2 \right)\\
&+\Pi\left( z\hat\theta x- \frac{\dot\beta }{\beta}z x -\frac{W^2}{2} \left({\hat\theta}-\frac{\dot\beta}{\beta}\right)^2 z^2
-\frac{1}{{2} } z^2 \right)\\
&+\Pi z\Delta_b\hat{\rho}\bar{u}\\
\leq&-kz^2-\Pi \kappa (x,\beta,\hat\theta)\Delta_b\hat{\rho}z^2.
\end{aligned}
\end{equation}
Then, substituting (\ref{u1}) into (\ref{d_hat_rho}) yields $\dot{\hat\rho}(t)=\gamma_\rho\Pi\text{sgn}(\ell_b)\kappa z^2$, where $\Pi>0$ and $\kappa(x,\beta,\hat\theta)>0$. When $b(t)>0$, according to Assumption 2, we can obtain  
$0<\ell_b<b(t)  $ and thus $\text{sgn}(\ell_b)>0 $ and $\Delta_b>0$,
implying that $ \dot{\hat\rho}(t)\geq0$. It follows from $\hat{\rho}(0)>0$ that $\hat\rho(t)>0$, and therefore    $z\Delta_b\hat\rho \bar u=-\kappa\Delta_b\hat\rho(t) z^2\leq 0$. Similarly, when $b(t)<0$,  according to Assumption 2, we can obtain $\Delta_b<0,~\text{sgn}(\ell_b)<0,~\dot{\hat\rho}(t)\leq0$ and therefore $z\Delta_b\hat\rho \bar u=-\kappa\Delta_b\hat\rho(t) z ^2\leq 0$ by selecting the initial condition $\hat{\rho}(0)<0$.  Recalling (\ref{d_hat_Theta}) and (\ref{d_hat_rho}), and noting the fact $-\Pi \kappa \Delta_b\hat{\rho}z^2\leq0$, it can be concluded that for any bounded initial $z(0)$, $V(t)\leq V(0)$,  which yields  $z(t),$ $\hat\theta,$  $\hat\rho$, and $W(x,\beta)$   are  bounded.

The boundedness of  $\Pi,$ $1/\Pi$,  $\Psi$ and $\kappa$ is   guaranteed by the boundedness of $z $ and $\beta(t)$,  it follows from (\ref{dz1-example-1}) and (\ref{dV-V})  that $\dot z \in\mathcal{L}_{\infty}$ and $z \in  \mathcal{L}_2$.  Therefore, invoking Barbalat's lemma one can conclude that $\lim_{t\rightarrow\infty}z (t)=0$, which further indicates that $\lim_{t\rightarrow\infty}x (t)=0$, hence the closed-loop system (\ref{scalar-system}) is asymptotically stable. Furthermore, by  using Lemma 1, we have  $x (t)\in\digamma_{\beta} =\left\{(t, x )\in \mathbb{R}_{\geq0}\times \mathbb{R}\big| |{x (t)}|/\psi^{-1}(\beta(t))<1\right\}$.

To show the asymptotic constancy of $\hat{\theta}$ and $\hat\rho$, recalling  (\ref{d_hat_Theta}), (\ref{d_hat_rho}), (\ref{dV-V}) and the fact that $z \in\mathcal{L}_2$, we have  $\dot{\hat{\theta}}\in\mathcal{L}_1$ and $ \dot{\hat\rho} \in\mathcal{L}_1$.   
Then, by using the argument similar to Theorem 3.1 in \cite{Krstic1996}, it is
concluded that $\hat{\theta}$ and $\hat\rho$ have a limit as $t\rightarrow\infty$. Furthermore,     it is seen from (\ref{d_hat_Theta}), (\ref{d_hat_rho}) and (\ref{u1}) that  the update laws $\dot{\hat{\theta}}\in\mathcal{L}_{\infty} $, $\dot{\hat{\rho}}\in\mathcal{L}_{\infty} $, and the control input $u=\hat\rho\bar u\in\mathcal{L}_{\infty} $.
This completes the proof.
$\hfill\blacksquare$

\section{Design for high-order time-varying systems}\label{section4}
Motivated by the design process for the first-order system, we now explore its applicability to more general higher-order system as described in (\ref{system}). For such strict-feedback system, we use  classical backstepping method \cite{krsticSCL}, with additional special treatment in each step, as detailed in what follows:

 \textit{Step 1:} Let $\alpha_1=x_2-z_2$ and according to (\ref{z1}), we can convert $\dot{x}_1=\phi_{1}^{\top}\theta(t)+x_2$ into the following $z_1$-dynamics 
\begin{equation}\label{dz1}
\begin{aligned}
\dot z_1  
=& \Pi\left(\alpha_1+z_2+\phi_{1 }^{\top}\theta(t) +\frac{\Psi }{\Pi}     \right)\\
=&\Pi\left(\alpha_1+z_2+\phi_{1 }^{\top}\hat\theta +\phi_{1 }^{\top}(\theta(t)-\ell_\theta) +\frac{\Psi }{\Pi}    \right.\\
&+\left.\phi_{1 }^{\top}( \ell_\theta-\hat\theta)  \right).
\end{aligned}
\end{equation}
where $\Pi(x_1,t)$ and $\Psi(x_1,t)$ are given below equation (\ref{dz1-example}), and $\ell_\theta\in\mathbb{R}^q$ is an unknown constant vector. By Hadamard's Lemma, one can express the regressor $\phi_1$ as $\phi_1(x_1)=\Phi_1(x_1)x_1$, where $\Phi_1(x_1)\in\mathbb{R}^q$ is smooth mapping. The third line of (\ref{dz1}) will be treated by the following tuning function 
\begin{equation} 
\tau_1(x_1,\beta )= \Gamma z_1 \Phi_{1}   \Pi x_1 
\end{equation} 
where $\Gamma=\Gamma^{\top}\in\mathbb{R}^{q\times q} $ is the positive adaptation gain.
Consider the Lyapunov function candidate 
\begin{equation}
V_1=\frac{1}{2}z_1^2+\frac{1}{2 }(\ell_\theta-\hat{\theta})^{\top}\Gamma^{-1}(\ell_\theta-\hat{\theta}),
\end{equation} 
then, by recalling Remark \ref{remark1}
\begin{equation}\label{dV1}
\begin{aligned}
\dot{V}_1=&\Pi\left(z_1\alpha_1 +z_1z_2+ \frac{\Psi_{x_1}}{\Pi}z_1 x_1 + z_1\phi_{1 }^{\top}\hat\theta  +z_1\phi_{1 }^{\top}\Delta_{\theta} 
\right)\\
&+(\ell_{\theta}-\hat\theta)^{\top}\Gamma^{-1}( \tau_1-\dot{\hat\theta}).
\end{aligned}
\end{equation} 
where $\Psi_{x_1}=\Psi/x_1$  is positive and invertible for all $x_1 \in \mathcal{L}_{\infty}$. Invoking Young's inequality, yields 
\begin{equation} \label{youngs1}
z_1\phi_{1 }^{\top}\Delta_{\theta}= z_1\Phi_{1 }^{\top}\Delta_{\theta}x_1 \leq \frac{\delta_{\Delta_{\theta}}}{2}\Phi_{1 }^{\top}\Phi_{1 }W_1^2z_1^2+\frac{\delta_{\Delta_{\theta}}}{2}z_1^2
\end{equation} 
where $W_1$ is shown  in Remark \ref{remark1}. The  virtual control law $\alpha_1$ is designed as
\begin{equation} \label{a1}
\begin{aligned}
\alpha_1( x_1,\bar\beta^{(1)},\hat\theta )=&-\frac{1}{\Pi}\left(k_1+\zeta_1\right)z_1-\frac{\Psi_{x_1}}{\Pi}  x_1-\phi_{1 }^{\top}\hat\theta   
\end{aligned}
\end{equation} 
where $\bar{\beta}^{(1)}=[\beta,\dot\beta]^\top$, $k_1>0$, and
\begin{equation*}
\zeta_1=\frac{1}{2}\left(\frac{1}{ \epsilon_{\psi}}+ \delta_{\Delta_\theta}\Pi\Phi_{1 }^{\top}\Phi_{1 }W_1^2+\Pi\delta_{\Delta_\theta}+(n-1)\delta_{\Delta_\theta}\right)
\end{equation*}
is the nonlinear damping gain with
$\epsilon_{\psi}>0$ and  $\delta_{\Delta_{\theta}}$ being the ``radius" of the compact set of  $\theta(t)$. $\Pi(x_1,t) \in\mathbb{R}_+$, $\Phi_{1}\in\mathbb{R}^{q}$ and $W_1\in{\mathbb{R}}$ are computable functions.  The resulting $\dot V_1$ is  
\begin{equation} \label{DV1}
\begin{aligned}
\dot V_1\leq& -k_1z_1^2+\Pi z_1z_2- \frac{(n-1)}{2}\delta_{\Delta_{\theta}}z_1^2- \frac{1}{2\epsilon_{\psi}}z_1^2\\
&+(\ell_{\theta}-\hat\theta)^{\top}\Gamma^{-1}( \tau_1-\dot{\hat\theta}) .
\end{aligned}
\end{equation} 
The second term  $\Pi z_1z_2$ in   the right hand side of (\ref{DV1}) can be canceled at the next step. 

\textit{Step 2}: Recall $\dot x_2=x_3+\phi_2^{\top}(\underline{x}_2)\theta(t)$ and let $\alpha_2=x_3-z_3$, we rewrite $\dot z_2=\dot x_2-\dot\alpha_1$ as
\begin{equation}\label{dz2}
\begin{aligned}
\dot{z}_2=&\alpha_2+z_3-\frac{\partial \alpha_1}{\partial x_1}x_2- \frac{\partial \alpha_1}{\partial \hat\theta }\dot{\hat\theta}   -\frac{\partial \alpha_1}{\partial \beta}\dot\beta -\frac{\partial \alpha_1}{\partial \dot\beta}\ddot\beta\\
&+\phi_2^{\top} \theta(t)-\frac{\partial \alpha_1}{\partial x_1}\phi_1^{\top}\theta(t).
\end{aligned}
\end{equation}
Define $w_2(\underline{x}_2,\hat\theta, \bar{\beta}^{(1)} )=\phi_2-\frac{\partial \alpha_1}{\partial x_1}\phi_1$,   then the second line of (\ref{dz2}) can be rewritten as
\begin{equation}\label{w}
\begin{aligned}
w_2^{\top}\theta(t)=w_2^{\top}\hat\theta+ w_2^{\top}(\theta(t)-\ell_{\theta})+w_2^{\top}( \ell_{\theta}-\hat\theta).
\end{aligned}
\end{equation}
Denote $\theta(t)-\ell_{\theta}$ by $\Delta_{\theta}$, and according to Assumption 1, there exist a known constant $\delta_{\Delta_{\theta}}$ such that $\delta_{\Delta_\theta}\geq |\Delta_{\theta}|$. Also note that
$z_1$,  and $\alpha_1(x_1,\bar{\beta}^{(1)},\hat\theta)$ are smooth and $ \alpha_1(0,\bar{\beta}^{(1)},\hat\theta)=0$,
and the $\hat\theta$- and $\bar{\beta}^{(1)}$-dependent change of coordinates between $\underline{z}_2$ and $\underline{x}_2$ is smooth, invertible, and $\underline{x}_2=0\Leftrightarrow\underline{z}_2=0$. Using Hadamard's Lemma, one can directly express $w_2$ as $w_2=W_2^{\top}(\underline{x}_2,\bar{\beta}^{(1)},\hat\theta)\underline{z}_2$, where $W_2(\underline{x}_2, \bar{\beta}^{(1)},\hat\theta  )\in\mathbb{R}^{2\times q}$ is a smooth mapping.
Therefore, one can calculate that
\begin{equation}\label{w2}
\begin{aligned}
z_2w_2^{\top}(\theta(t)-\ell_{\theta})=&z_2\Delta_{\theta}^{\top}w_2=z_2\Delta_{\theta}^{\top}W_2^{\top}\underline{z}_2\\
\leq&\frac{1}{2}\delta_{\Delta_\theta}|W_2|_{_{\mathsf{F}}}^2z_2^2+\frac{\delta_{\Delta_\theta}}{2}\underline{z}_2^{\top}\underline{z}_2\\
=&\frac{\delta_{\Delta_\theta}}{2}\left(|W_2|_{_{\mathsf{F}}}^2+1\right)z_2^2+\frac{\delta_{\Delta_\theta}}{2} {z}_1^2
\end{aligned}
\end{equation}
where $\underline{z}_2^{\top}\underline{z}_2=z_1^2+z_2^2$ is used and $|W_2|_{_{\mathsf{F}}}=\sqrt{\sum_{i=1}^{2}\sum_{j=1}^{q}({W_2}_{ij})^2}$ denotes the Frobenius norm.
Choosing the Lyapunov function candidate $V_2=V_1+\frac{1}{2}z_2^2$, its  derivative  along the trajectories of (\ref{system}) is
\begin{equation}\label{dV2}
\begin{aligned}
\dot V_2\leq&-k_1z_1^2+\Pi z_1z_2- \frac{(n-1)}{2}\delta_{\Delta_{\theta}}z_1^2- \frac{1}{2\epsilon_{\psi}}z_1^2\\
&+z_2\alpha_2+z_2\left(-\frac{\partial \alpha_1}{\partial x_1}x_2 -\frac{\partial \alpha_1}{\partial \hat\theta }\dot{\hat\theta}-\frac{\partial \alpha_1}{\partial \beta}\dot\beta\right)\\
&+\frac{\delta_{\Delta_\theta}}{2}\left(|W_2|_{_{\mathsf{F}}}^2+1\right)z_2^2+\frac{\delta_{\Delta_\theta}}{2} {z}_1^2+z_2w_2^{\top}\hat\theta+z_2z_3\\
& +  (\ell_{\theta}-\hat\theta)^{\top}\Gamma^{-1}(\Gamma w_2z_2+\tau_1-\dot{\hat\theta}) .
\end{aligned} 
\end{equation}
According to (\ref{dV2}), we  design the tuning function as
\begin{equation}\label{theta2}
\tau_2(\underline{x}_2,\bar{\beta}^{(1)},\hat\theta)=\tau_1+\Gamma w_2z_2.
\end{equation}
In addition, the virtual control law $\alpha_2$ is constructed as
\begin{equation}\label{a2}
\begin{aligned}
\alpha_2(\underline{x}_2,\bar{\beta}^{(2)},\hat{\theta} )=&- \Pi z_1  -(k_2+\zeta_2)z_2-w_2^{\top}\hat\theta\\
&+ \frac{\partial \alpha_1}{\partial x_1}x_2+\frac{\partial \alpha_1}{\partial \beta}\dot\beta+\frac{\partial \alpha_1}{\partial \dot\beta}\ddot\beta  +\frac{\partial \alpha_1}{\partial \hat\theta }\tau_2 
\end{aligned}
\end{equation}
where $\bar\beta^{(2)}=[\beta,\dot\beta,\ddot{\beta}]^\top$, $k_2>0$,  and $\zeta_2(\underline{x}_2,\bar\beta^{(1)},\hat{\theta})$ is the nonlinear damping gain, as follows
\begin{equation}\label{zeta2}
\begin{aligned}
\zeta_2=\frac{1}{2}\left(\delta_{\Delta_\theta}|W_2|_{_{\mathsf{F}}}^2+(n-1)\delta_{\Delta_\theta}+\frac{1}{\epsilon_{\psi}}\right).
\end{aligned}
\end{equation}
After some simplifications and using (\ref{theta2}) and (\ref{a2}),  we  express  (\ref{dV2})   as
\begin{equation}\label{dV2-1}
\begin{aligned}
\dot V_2\leq&-k_1z_1^2-k_2z_2^2  -\frac{1}{2}\left((n-2){\delta_{\Delta_\theta}} +\frac{1}{\epsilon_{\psi}}\right)\underline{z}_2^{\top}{\underline{z}_2}\\
&+z_2z_3 +\left(z_2\frac{\partial \alpha_1}{\partial \hat\theta}+(\ell_{\theta}-\hat\theta)^{\top}\Gamma^{-1}\right)( \tau_2-\dot{\hat\theta})
\end{aligned}
\end{equation}
where $z_2z_3$ can be canceled at the next step.

\textit{Step 3}: Introducing $\alpha_3=x_4-z_4$ and according to $z_3=x_3-\alpha_2$,  we can transform $\dot x_3=x_4+\phi_3^{\top}(\underline{x}_3)\theta(t)$  to the following $z_3$-dynamics  
\begin{equation}\label{dz3}
\begin{aligned}
\dot{z}_3=&\alpha_3+z_4- \frac{\partial \alpha_2}{\partial x_1}x_{2}-\frac{\partial \alpha_2}{\partial x_2}x_{3}- \frac{\partial \alpha_2}{\partial \beta }\dot\beta- \frac{\partial \alpha_2}{\partial\dot \beta }\ddot\beta \\
&- \frac{\partial \alpha_2}{\partial \ddot\beta }\dddot\beta -\frac{\partial \alpha_2}{\partial \hat\theta }\dot{\hat\theta}+w_3^{\top}\hat\theta  +w_3^{\top}(\ell_{\theta}- \hat{\theta})\\
&+w_3^{\top}( {\theta}(t)-\ell_{\theta}).
\end{aligned} 
\end{equation} 
Now we choose the Lyapunov function candidate $V_3=V_2+\frac{1}{2}z_3^2$, then
\begin{equation}\label{dV3}
\begin{aligned}
\dot V_3\leq &-k_1z_1^2-k_2z_2^2  -\frac{1}{2}\left((n-2){\delta_{\Delta_\theta}} +\frac{1}{\epsilon_{\psi}}\right)\underline{z}_2^{\top}{\underline{z}_2}\\
& +z_3z_4+ \left(z_2\frac{\partial \alpha_1}{\partial \hat\theta}+(\ell_{\theta}-\hat\theta)^{\top}\Gamma^{-1}\right)( \tau_2-\dot{\hat\theta})\\
&-z_3 \left(\sum_{j=1}^{2}\frac{\partial \alpha_2}{\partial x_j}x_{j+1}+\sum_{j=0}^{2}\frac{\partial \alpha_2}{\partial \beta^{(j)}}\beta^{(j+1)}+\frac{\partial \alpha_2}{\partial \hat\theta }\dot{\hat\theta}\right)\\
& +z_3\alpha_3+z_2z_3+z_3w_3^{\top}\hat\theta+z_3w_3^{\top}(\ell_{\theta}- \hat{\theta})\\
&+z_3w_3^{\top}(\theta(t)-\ell_{\theta}) 
\end{aligned} 
\end{equation}
where $w_3(\underline{x}_3,\bar\beta^{(2)}, \hat\theta)=\phi_3-\frac{\partial \alpha_2}{\partial x_1}\phi_1-\frac{\partial \alpha_2}{\partial x_2}\phi_2\in\mathbb{R}^q$ is the new regressor vector, and it can be verified that $w_3(0,\bar\beta^{(2)},\hat\theta)=0$.  
Using the analysis similar to that used in (\ref{w})-(\ref{w2}), one can express   $w_3$ as $w_3=W_3^{\top}(\underline{x}_3,\bar\beta^{(2)},\hat\theta)\underline{z}_3$,  where $W_3\in\mathbb{R}^{3\times q}$ is a smooth mapping.  Therefore,   we obtain an upper bound of the last line of (\ref{dV3}), as follows
\begin{equation}\label{w3}
\begin{aligned}
z_3w_3^{\top}(\theta(t)-\ell_{\theta}) 
\leq \frac{\delta_{\Delta_\theta}}{2}\left(|W_3|_{_{\mathsf{F}}}^2+1\right)z_3^2+\frac{\delta_{\Delta_\theta}}{2} \underline{z}_2^{\top}\underline{z}_2.
\end{aligned}
\end{equation}
Then, we design the following tuning function and virtual control law, respectively
\begin{equation}\label{tau3}
\tau_3(\underline{x}_3,\bar\beta^{(2)},\hat\theta)=\tau_2+\Gamma w_3z_3
\end{equation}
\begin{equation}\label{alpha3}
\begin{aligned}
\alpha_3&(\underline{x}_3,\bar\beta^{(3)},\hat\theta)=-z_2-(k_3+\zeta_3)z_3-w_3^{\top}\hat\theta +\frac{\partial \alpha_2}{\partial \hat\theta }\tau_3\\
&+\sum_{j=1}^{2}\frac{\partial \alpha_2}{\partial x_j}x_{j+1}+\sum_{j=0}^{2}\frac{\partial \alpha_2}{\partial \beta^{(j)}}\beta^{(j+1)} +\frac{\partial \alpha_1}{\partial\hat\theta}\Gamma z_2w_3
\end{aligned} 
\end{equation}
where $\bar\beta^{(3)}=[\beta,\dot\beta,\ddot{\beta},\dddot{\beta}]^\top$, $k_3>0$,  and  
\begin{equation}\label{zeta3}
\zeta_3 =\frac{1}{2}\left(\delta_{\Delta_\theta}|W_3|_{_{\mathsf{F}}}^2+(n-2)\delta_{\Delta_\theta}+\frac{1}{\epsilon_{\psi}}\right).
\end{equation}
Now,  in virtue of (\ref{tau3}) and (\ref{alpha3}), we can rewrite $\dot V_3$ as 
\begin{equation}
\begin{aligned}
\dot V_3\leq& -\sum_{j=1}^{3}k_jz_j^2+z_3z_4-\frac{1}{2}\left((n-3)\delta_{\Delta_\theta}+\frac{1}{\epsilon_{\psi}}\right)\underline{z}_3^{\top}\underline{z}_3\\
&+\left(z_2\frac{\partial\alpha_1}{\partial \hat\theta}+z_3\frac{\partial \alpha_2}{\partial \hat\theta}+(\ell_{\theta}-\hat{\theta})^{\top}\Gamma^{-1}\right)( \tau_3-\dot{\hat\theta}).
\end{aligned} 
\end{equation}
where $z_3z_4$ can be canceled at the next step.

\textit{Step $i~(i=3,\cdots,n-1)$}: We are now in the position to summarize the expression of the input signals by  previous design steps.  
\begin{equation}\label{ii}
\begin{cases}	
z_i=x_i-\alpha_{i-1},\\
w_i(\underline{x}_i,\bar\beta^{(i-1)},\hat\theta)=\phi_i-\sum_{j=1}^{i-1}\frac{\partial \alpha_{i-1}}{\partial x_j}\phi_{j },\\ 
\tau_i(\underline{x}_i,\bar\beta^{(i-1)},\hat\theta)=\tau_{i-1}+\gamma_{\theta}w_iz_i,\\
\alpha_i(\underline{x}_i,\bar\beta^{(i )},\hat\theta)= -z_{i-1}-(k_i+\zeta_i)z_i-w_i^{\top}\hat\theta\\
~~~~~~~ +\sum_{j=1}^{i-1}\frac{\partial \alpha_{i-1}}{\partial x_j}x_{j+1}+\sum_{j=0}^{i-1}\frac{\partial \alpha_{i-1}}{\partial \beta^{(j)}}\beta^{(j+1)}\\
~~~~~~~  +\sum_{j=2}^{i-1}\frac{\partial \alpha_{j-1}}{\partial \hat\theta }\Gamma z_jw_i+ \frac{\partial \alpha_{i-1}}{\partial \hat\theta }\tau_i ,\\
\zeta_i =\frac{1}{2}\left(\delta_{\Delta_\theta}|W_i|_{_{\mathsf{F}}}^2+(n+1-i)\delta_{\Delta_\theta}+\frac{1}{\epsilon_{\psi}}\right) ,
\end{cases}  
\end{equation}
where $k_i>0$, $\epsilon_{\psi}>0$, $\bar{\beta}^{(i)}=[\beta,\dot\beta,\cdots,\beta^{(i)}]^\top\in\mathbb{R}^{i+1}$, and $W_i\in\mathbb{R}^{i\times q}$ is a smooth mapping.  Based upon (\ref{ii}),
the  derivative of $V_i=V_{i-1}+\frac{1}{2}z_i^2$ can be computed as 
\begin{equation}
\begin{aligned}
\dot V_i\leq& -\sum_{j=1}^{i}k_jz_j^2+z_iz_{i+1}-\frac{1}{2}\left((n-i)\delta_{\Delta_\theta}+\frac{1}{\epsilon_{\psi}}\right)\underline{z}_i^{\top}\underline{z}_i\\
&+\Big(\sum_{j=1}^{i-1}\frac{\partial \alpha_j}{\partial \hat\theta}z_{j+1}+(\ell_{\theta}-\hat{\theta})^{\top}\Gamma^{-1}\Big)( \tau_i-\dot{\hat\theta}).
\end{aligned}
\end{equation}

\textit{Step $n$}: This step is different from the previous steps. On one hand, the actual control law and update law of $\hat\theta$   should be designed in this step. On the other hand, we need to extend the \textit{congelation of variables} for time-varying parameters in the feedback path to the scenario that time-varying parameters in the input path. 

To proceed, we rewrite $\dot x_n=\phi_{n}^{\top}\theta(t)+b(t)u$ as
\begin{equation}\label{dz}
\begin{aligned}
\dot z_n
=&w_n^{\top}\theta(t)+b(t)u- \frac{\partial \alpha_{n-1}}{\partial \hat\theta }\dot{\hat{\theta}}- \sum_{j=1}^{n-1}\frac{\partial \alpha_{n-1}}{\partial x_j}x_{j+1}\\
& -\sum_{j=0}^{n-1}\frac{\partial \alpha_{n-1}}{\partial \beta^{(j)}}\beta^{(j+1)}
\end{aligned}
\end{equation}
where $w_n=\phi_{n}-\sum_{j=1}^{n-1}\frac{\partial \alpha_{i-1}}{\partial x_j}\phi_j$. The main different will start from the following design. For the next developments we need the following intermediate result by means of   $u=\hat\rho\bar u$ 
\begin{equation}\label{39}
  \begin{aligned}
	z_n\dot z_n  
	=&z_nw_n^{\top}\hat\theta+z_nw_n^{\top} (\theta(t)-\ell_\theta)+z_nw_n^{\top} ( \ell_\theta-\hat\theta)\\ 
	&+ z_n\bar u+z_n(b(t)-\ell_b)\hat\rho\bar u+z_n\ell_b\left(\frac{1}{\ell_b}-\hat{\rho}\right)\bar u\\
	&-z_n \frac{\partial \alpha_{n-1}}{\partial \hat\theta }\dot{\hat{\theta}}-z_n\sum_{j=1}^{n-1}\frac{\partial \alpha_{n-1}}{\partial x_j}x_{j+1}\\
	&-z_n\sum_{j=0}^{n-1}\frac{\partial \alpha_{n-1}}{\partial \beta^{(j)}}\beta^{(j+1)} .
	\end{aligned}  
\end{equation}
where  $\ell_b$ is an unknown constant  which can be regard  as the average of $b(t)$, $\hat\rho$ is an ``estimate" of $1/\ell_b$ and denote $b(t)-\ell_b$ by $\Delta_b$. Note that we need  $\delta_{\Delta_{\theta}}$ to construct the nonlinear damping gain to cancel the effect of unknown $\theta(t)$, as  our previous steps do. However, the same method cannot be used directly for dealing with $b(t)$ since the perturbation term $z_1\Delta_{b}\hat{\rho}\bar{u}$ is coupled with the control input. 
Here we apply a special way to cope with the unknown time-varying quantities, \textit{i.e.}, designing $\bar u$ skillfully to ensure the perturbation term $z_n(b(t)-\ell_b)\hat\rho\bar u$ in the second of (\ref{39}) is always negative. 

Consider the Lyapunov function candidate 
\begin{equation}
\begin{aligned}
V_n= V_{n-1}+\frac{|\ell_b|}{2\gamma_\rho}\left(\frac{1}{\ell_b}-\hat\rho\right)^2 
\end{aligned}
\end{equation}
then,
\begin{equation}\label{dVn}
\small{\begin{aligned}
	\dot V_n=&\dot V_{n-1}-\frac{|\ell_b|}{ \gamma_\rho}\left(\frac{1}{\ell_b}-\hat\rho\right)\dot{\hat\rho} \\
	\leq&-\sum_{j=1}^{n-1}k_jz_j^2-\frac{1}{2}\left( \delta_{\Delta_\theta}+\frac{1}{\epsilon_{\psi}}\right)\underline{z}_{n-1}^{\top}\underline{z}_{n-1}\\
	&+\left(\sum_{j=1}^{n-1}\frac{\partial \alpha_j}{\partial \hat\theta}z_{j+1}+(\ell_{\theta}-\hat{\theta})^{\top}\Gamma^{-1}\right)( \tau_{n-1}-\dot{\hat\theta})\\
	&+z_{n-1}z_n+z_nw_n^{\top}\hat\theta+ z_nw_n^{\top}\Delta_{\theta}+z_nw_n^{\top} ( \ell_\theta-\hat\theta)\\
	&+z_n\bar u+z_n\Delta_{b}\hat\rho\bar u-z_n\sum_{j=0}^{n-1}\frac{\partial \alpha_{n-1}}{\partial \beta^{(j)}}\beta^{(j+1)}\\
	&-z_n \sum_{j=1}^{n-1}\frac{\partial \alpha_{n-1}}{\partial x_j}x_{j+1}-z_n\frac{\partial \alpha_{n-1}}{\partial \hat\theta }\dot{\hat{\theta}}\\ 
	& +z_n\ell_b\left(\frac{1}{\ell_b}-\hat{\rho}\right)\bar u-\frac{|\ell_b|}{ \gamma_\rho}\left(\frac{1}{\ell_b}-\hat\rho\right)\dot{\hat\rho}  
	\end{aligned}}
\end{equation} 
where $w_n =\phi_n-\sum_{j=1}^{n-1}\frac{\partial \alpha_{n-1}}{\partial x_j}\phi_{j }$. Now, to cancel the third and last lines of (\ref{dVn}), we design the update laws for the parameters $\hat\theta$ and $\hat\rho$, as follows
\begin{equation}\label{42}
\begin{aligned}
\dot{\hat\theta}=&\tau_n=\tau_{n-1}+\Gamma  w_nz_n\\
=&\Gamma \Big(z_1\Phi_1 \Pi x_1 +\sum_{j=2}^{n}w_jz_j\Big),  
\end{aligned}
\end{equation}
\begin{equation}\label{48}
\dot{\hat{\rho}}=-\gamma_{\rho}\text{sgn}(\ell_b)z_n\bar u.
\end{equation}  
\begin{remark}\label{remark2}
	Define $\Omega(\underline{x}_{n},\bar{\beta}^{(n )},\hat\theta)=z_{n-1}+w_n^{\top}\hat\theta- \frac{\partial \alpha_{n-1}}{\partial \hat\theta }\tau_n-\sum_{j=1}^{n-1}\frac{\partial \alpha_{n-1}}{\partial x_j}x_{j+1}- \sum_{j=0}^{n-1}\frac{\partial \alpha_{n-1}}{\partial \beta^{(j)}}\beta^{(j+1)} -\sum_{j=2}^{n-1}\frac{\partial \alpha_{j-1}}{\partial \hat\theta}\Gamma z_jw_n$.  
	It can be further verified, for $i=1,\cdots,n-1$,  that  $\alpha_i$, $w_i$, $\tau_i$ and $\Omega$ are smooth, and $\alpha_i=w_i=\tau_i=\Omega=0$ if $\underline{x}_i=0$.
	Note also that the coordinate transformation 
	\begin{equation}
	z_1=\frac{\beta(t) \psi(x_1)}{\beta^2(t)-\psi^2(x_1)}
	\end{equation} 
	and $z_i=x_i-\alpha_{i-1}$ ($i>1$) is also smooth, invertible and $x_i=0\Leftrightarrow z_i=0$. 
	According to Hadamard's Lemma,   $w_n(x,\bar\beta^{(n-1)},\hat\theta)$ and $\Omega$  can be expressed as $w_n=W_n^{\top}\mathbf{z}$
	and   $\Omega=\bar\Omega^{\top} \mathbf{z} $, respectively,
	with $\mathbf{z}=[z_1,\cdots,z_n]^{\top}$, $W_n\in\mathbb{R}^{n\times q}$ and $\bar\Omega\in\mathbb{R}^{ n}$ being smooth mappings.
\end{remark} 

Applying Young's inequality with $\epsilon_{\psi}>0$, yields
\begin{equation*}
\begin{aligned}
&z_n\Omega= z_n\bar\Omega^{\top} \mathbf{z}\leq\frac{1}{2}\left(\epsilon_{\Omega}|\bar\Omega|^2+\frac{1}{\epsilon_{\Omega}}\right)z_n^2+\frac{1}{2\epsilon_{\Omega}}\underline{z}_{n-1}^{\top}\underline{z}_{n-1},
\end{aligned}
\end{equation*}
\begin{equation*}
\begin{aligned}
z_nw_n^{\top}\Delta_{\theta} \leq \frac{\delta_{\Delta_\theta}}{2}\left(|W_n|_{_{\mathsf{F}}}^2+1\right)z_n^2+\frac{\delta_{\Delta_\theta}}{2} \underline{z}_{n-1}^{\top}\underline{z}_{n-1}.
\end{aligned}
\end{equation*}
Finally, we choose the actual control law  $u=\hat\rho\bar u$ such that the time-varying perturbed term $z_n\Delta_{b}\hat\rho\bar u$  is nonpositive
\begin{equation}\label{45}
\left\{\begin{array}{lr}
\bar u=-\kappa\left(x,\beta,\cdots,\beta^{(n )},\hat{\theta}\right)z_n\\
\kappa=k_n+\frac{1}{2}\left(\delta_{\Delta_\theta}|W_n|_{_{\mathsf{F}}}^2+\delta_{\Delta_\theta}+\frac{1}{\epsilon_{\Omega}}+ \epsilon_{\Omega}|\bar\Omega|^{2} \right) 
\end{array}\right.
\end{equation}
where $k_n>0$. Inserting (\ref{42})-(\ref{45}) into (\ref{dVn}), yields
\begin{equation}\label{Vn}
\dot V_n\leq -\sum_{j=1}^{n}k_jz_j^2- \kappa \Delta_{b}\hat\rho(t) z_n^2.
\end{equation}

\section{Stability analysis}\label{section5}
Firstly, it can be shown that $\hat\rho(t)$ in the right hand side of (\ref{Vn}) is a monotonic increasing (or decreasing) function by calculating equation (\ref{48}) as $\dot{\hat\rho}= \gamma_{\rho}\text{sgn}(\ell_b)\kappa z_n^2$. 
In addition, one can select $\hat\rho(0)>0$ when $0<\ell_b\leq b(t)$ (in this case, $\Delta_b>0$) to make sure that $\hat\rho(t)>0$, thereby obtaining $ -\kappa\Delta_{b}\hat{\rho}z_n^2<0$. 
Similarly, one can select $\hat\rho(0)<0$ when $ b(t) \leq\ell_b<0$ (in this case, $\Delta_b<0$) to make sure that $\hat\rho(t)<0$, thereby obtaining $z_n\Delta_{b}\hat\rho\bar u=-\kappa\Delta_{b}\hat{\rho}z_n^2\leq0$ again. 
Therefore, formula (\ref{Vn}) can be simplified as $\dot V_n\leq-\sum_{j=1}^{n}k_jz_j^2\leq 0$, which guarantees that $z$,  $\hat\theta$, and $\hat{\rho}$ are bounded for all $t\geq 0.$

Next, in view of  Remark  \ref{remark1} and the boundedness of $z$, it follows that  $W_1$, $1/\Pi$ and $\Pi $ are bounded,  and therefore   $\tau_1$ and $\alpha_1$ are bounded, which further proves the boundedness of $x_2$ along with the coordinate transformation $x_2=z_2+\alpha_1$ and the boundedness of $w_2$ due to (\ref{ii}). Hence  $W_2$, $\tau_2$ and $\alpha_2$ are also bounded. Following this line of argument, the boundedness of state $x_i$, virtual control $\alpha_i$ $(i=3, \cdots, n-1)$, and the actual control input $u$ are ensured. In addition, it is seen from (\ref{42})  and (\ref{48}) that  $\dot{\hat{\theta}}\in\mathcal{L}_{\infty} $ and $\dot{\hat{\rho}}\in\mathcal{L}_{\infty} $. To show the asymptotic constancy of $\hat{\theta}$ and $\hat\rho$,  it follows from $\dot V_n\leq -\sum_{j=1}^{n}k_jz_j^2$ that $z_j\in\mathcal{L}_2$,   then from (\ref{42}) and (\ref{48}) we get $\hat{\theta}\in\mathcal{L}_1$ and $\hat\rho\in\mathcal{L}_1$; 
by using the argument similar to Theorem 3.1 in \cite{Krstic1996}, it is
concluded that $\hat{\theta}$ and $\hat\rho$ have a limit as $t\rightarrow\infty$. 

Finally,  it follows from (\ref{dz1}), (\ref{dz2}), (\ref{dz3}), (\ref{dz}) and (\ref{Vn}) that $\dot {\mathbf{z}}\in\mathcal{L}_{\infty}$ and $\mathbf{z}\in\mathcal{L}_{2}\cap\mathcal{L}_{\infty}$, then using  Barbalat's  Lemma yields $\lim_{t\rightarrow\infty}\mathbf{z}(t)=0$, which further indicates that $\lim_{t\rightarrow\infty}x(t)=0$.  Therefore, the closed-loop system is   asymptotically stable.   By  virtue of Lemma 1, we get  the output $x_1(t)$ is always constrained within the prescribed performance funnel $\digamma_{\beta}:=\left\{(t, x_1)\in[0,\infty)\times \mathbb{R}\big| |{x_1(t)}|/\psi^{-1}(\beta(t))<1\right\}$.

The above facts prove the following result:

\begin{theorem}
	Suppose  that the design procedure is applied to the nonlinear system (\ref{system}) with time-varying parameters. Then,  
	the closed-loop system is asymptotically stable and the system output  $x_1(t)$ is always confined within the prescribed performance funnel $\digamma_{\beta}:=\left\{(t, x_1)\in\mathbb{R}_{\geq0}\times \mathbb{R}\big| |{x_1(t)}|/\psi^{-1}(\beta(t))<1\right\}$ and ultimately decays to zero. Furthermore,   $\lim_{t\rightarrow\infty}\hat{\theta}$ and $\lim_{t\rightarrow\infty}\hat{\rho}$  exist  but they are not necessarily 	equal to $\ell_\theta$ and $1/\ell_b$. In addition, the control input and update laws remain uniformly bounded over $[0,\infty)$.     
	$\hfill\blacksquare$
\end{theorem}

\begin{remark}\label{remarkIV-2}
	The  proposed controller primarily consists of three units: robust unit, $\theta(t)$-adaptive unit and $b(t)$-adaptive unit.  Note that the $\theta(t)$-adaptive unit  is completely equivalent to the design of update laws in classical adaptive control since  we use the unknown constant $\ell_{\theta}$    to replace $\theta(t)$. The time-varying perturbation term $\Delta_{\theta}(t)$ caused by $\theta(t)-\ell_\theta$ is allocated to the robust unit for processing.  This is an easy-to-understand and easy-to-implement solution, in other words, the proposed controller is simple in structure and user-friendly in design.
	In addition, $b(t)$-adaptive unit is deliberately designed for unknown and time-varying control gain, whose main purpose is to ensure the perturbation term $z_n\Delta_{b}\hat\rho\bar u\leq 0$, thereby avoiding the adaptive parameter drifting caused by unknown gains. 
\end{remark}

\begin{remark}\label{remarkIV-1}
	The control scheme involves the selection of $\{k_i\}_{i=1}^n>0$,
	$\hat{\theta}(0)\geq 0$, $\hat{\rho}(0)>0$, $\delta_{\Delta_\theta}>0$,  $\epsilon_{\psi}>0$, and $\Gamma>\mathbf{0}$, which theoretically can be chosen quite
	arbitrarily by  users. Certain compromise between convergence rate and control effort needs to be made when making the selection for those parameters for a given system. For example,  the parameters $k_i$ and $\delta_{\Delta_\theta}$ are proportional to convergence rate and control effort in this paper, and thus   reducing the input effort will cause the convergence rate to slow down. However, it is worth noting that   the prescribed constraint rule will not be  violated no matter how the parameters are selected. 
\end{remark}
\begin{remark}  
	Compared with the previous work \cite{Zhao} on adaptive exponential regulation for systems with time-invariant parameters, the proposed method provides a simpler solution, and without loss of final control accuracy, completely eliminates the necessity for the control gain to grow with time ceaselessly. 
\end{remark}
\begin{remark} 
	Different from traditional guaranteed performance control  (see, for instance, \cite{Benchlioulis,zhang,zhangcyber-global}) that can only achieve bounded regulation and the size of the regulation residual set is reversely proportional to the control gain,  such that higher final control precision is essentially at the price of large control gain, the proposed control method  is able to steer each system state to zero asymptotically without the need for prohibitively large controller gain. Furthermore, no matter how small the control gain is, $-\psi^{-1}(\beta)<x_1(t)<\psi^{-1}(\beta)$ always holds.
\end{remark}
\begin{remark} 
	Our control scheme benefits from  Chen \& Astolfi’  Method \cite{chenkaiwen,chenkaiwenACC,chenkaiwenIFAC} in dealing with unknown  time-varying parameters, furthermore, by introducing the performance function and employing a novel coordinate transformation, our control scheme is able to explicitly address  global transient behavior  of system output, together with its steady-state performance.
\end{remark}

\section{simulation}
To verify the effectiveness of the proposed control method,  we consider the following system\footnote{Note that  when $\theta(t)$ is an unknown constant and $b(t)=1$, this model is a simplified  version of the one studied by  \cite{Zhao}, where the exponential regulation is proposed for a class of strict-feedback systems with known control gain and unknown constants $\theta$.}
\begin{equation}  
\begin{array}{l}
\dot{x}_1 =\theta(t)x_1 +x_2;~~~~ \\
\dot{x}_2 =b(t)u  ;~~~~\\y(t)=x_1
\end{array}
\end{equation} 
with  fast time-varying parameters\footnote{Here $b(t)$ and $\theta(t)$ are fast time-varying parameters and they are only piecewise continuous yet $b(t)$ may undergo sudden changes. Therefore, some classical adaptive schemes  \cite{Goodwin-nonPE,Marino} are not available because those methods require the  parameters be slow time-varying (\textit{i.e.,} here exists a parameter $\epsilon$ such that $|\dot{\theta}(t)|<\epsilon$ and $|\dot b(t)|<\epsilon$).} 
\begin{equation}
b(t)=2+0.1\cos(x_1)+\text{sign}({x_1x_2} )
\end{equation}
\begin{equation}
\theta(t)=2 + 0.8\sin(t) + \sin(x_1x_2) + 0.2\sin (x_1 t)  +\text{sign}(\sin(t)).
\end{equation} 
It is not difficult to verify that Assumptions 1-2 are satisfied. 
The control objective is to make the state $x_1$ moves back to zero at a prescribed rate no slower than exponential and ultimately converges to zero. 
Now we consider three controllers:  Controller 1 is the adaptive controller proposed by \textit{Chen \& Astolfi} in \cite{chenkaiwen}; Controller 2 is the semi-global adaptive prescribed performance controller which can be obtained by combining Controller 1 and the controller proposed in \cite{Benchlioulis}; Controller 3 is the global controller proposed in Theorem 2.  In fact, Controller 2 can be viewed as a special case of Controller 3. 
For fair comparison, we set $[x_1(0);x_2(0)]=[1;-1]$, $k_1=k_2=\gamma_{\rho}=0.1$,  $\delta_{\Delta_\theta}=1,$  $\Gamma=0.1I$, $\hat{\theta}(0)=0$ and  $\hat\rho(0)=0.25$ for all controllers. In addition, we select $\beta(t)=4e^{-0.4t}+0.1$ and $z_1=\tan(\pi x_1/(2\beta))$ for Controller 2, and select $\beta_1(t)=0.9e^{-0.4t}+0.1$  for Controller 3. 

The responses of the state signals are shown in Figs 1-2, and the   responses of control input signals are shown in Fig 3. The evolution of adaptive parameters $\hat{\theta}$ and $\hat{\rho}$ are shown in Fig. 4 and 5, respectively. In addition, we also  illustrate the time-varying parameters $\theta(t)$ and $b(t)$ in Fig 6, which shows that the  state-dependent parameters are fast time-varying and nondifferentiable. From these simulation results, we know that the proposed controllers outperforms the adaptive controller in \cite{chenkaiwen}, since the transient behavior of the system can be confined to a prescribed performance boundary.  In particular, compared Controller 1 with Controllers 2-3, one can find  a counterintuitive phenomenon, that is, based on the previous parameter selection, faster system response   can be achieved without an increase in control effort. In short, all results show that the proposed methods are powerful enough to stabilize the nonlinear system with fast time-varying parameters.

\begin{figure}[!htbp]  \label{figV1} 
	\begin{center}
		\includegraphics[height=4.8 cm]{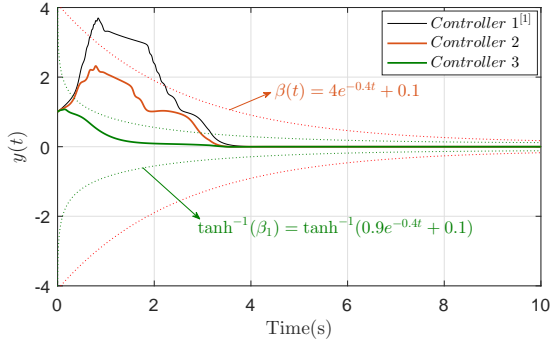} 
		\caption{ The evolution of $y(t)$. }
	\end{center}
\end{figure}
\begin{figure}[!htbp]  \label{figV2} 
	\begin{center}
		\includegraphics[height=4.8  cm]{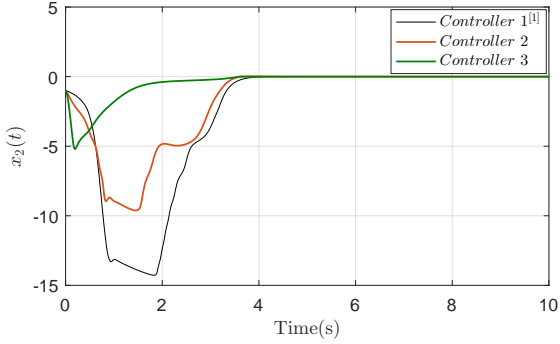}
		\caption{ The evolution of $x_2(t)$. }
	\end{center}
\end{figure}
\begin{figure}[!htbp]  \label{figV3} 
	\begin{center}
		\includegraphics[height=4.8   cm]{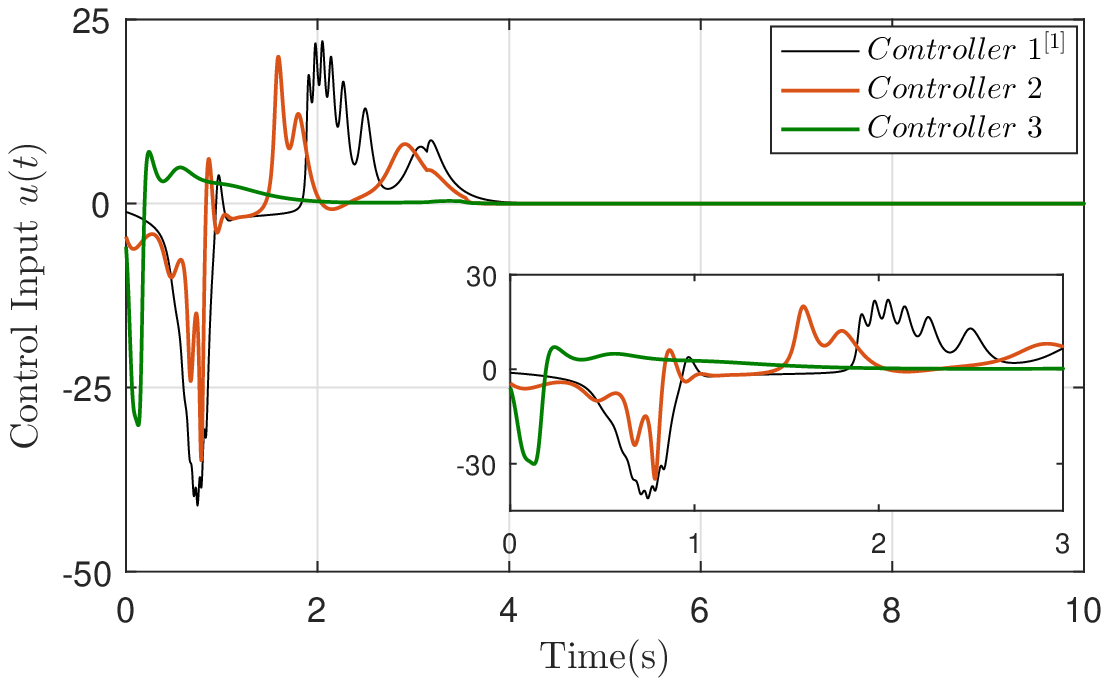}
		\caption{ The evolution of $u(t)$. }
	\end{center}
\end{figure}
\begin{figure}[!htbp]  \label{figV4} 
	\begin{center}
		\includegraphics[height=4.8  cm]{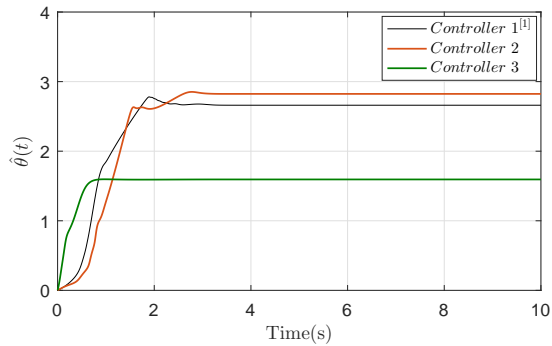}
		\caption{ The evolution of   $\hat{\theta}(t)$. }
	\end{center}
\end{figure}
\begin{figure}[!htbp]  \label{figV5} 
	\begin{center}
		\vspace{0.9em}\includegraphics[height=4.5 cm]{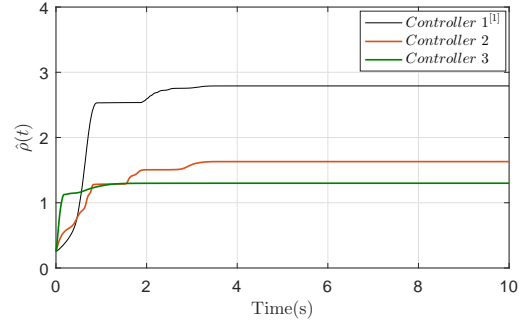}
		\caption{ The evolution of    $\hat{\rho}(t)$. }
	\end{center}
\end{figure}
\begin{figure}[!htbp]   \label{figV6} 
	\begin{center}
		\vspace{0.7em}\includegraphics[height=4.5 cm]{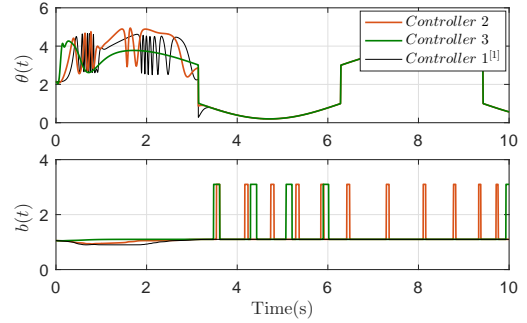}
		\caption{System time-varying parameters $\theta(t)$ and $b(t)$. }
	\end{center}
\end{figure}

\section{conclusion}
This work  presents an adaptive control strategy with guaranteed performance for strict-feedback nonlinear systems involving fast time-varying parameters. It is shown that with this strategy,  not only each system state is regulated to zero asymptotically, but also the system output is strictly confined within an exponentially decaying boundary, making system output well behaved during transient period and steady-state phase. We start with a simple scalar system with time-varying parameters in the feedback path and input path to illustrate our core idea in addressing time-varying parameters and output performance constraint simultaneously. By using classical Backstepping technology and nonlinear damping, we then extend our method to higher-order system and remove the need for overparametrization. Furthermore, the diversity of performance function selection and the diversity of normalized function selection  together with the independence on initial conditions imply the universal of our controller, and  simulation comparisons confirm the effectiveness and benefits of these methods.

Prior to the work, the prevailing wisdom in adaptive control in the context of  exponential stability for time-varying systems is that certain persistent excitation conditions (sufficiently rich signals) must be present. Here in this work we develop a method that  achieves exponential convergence, pointwise in time, without the need for PE conditions. An interesting future research topic is to study the exponential stabilization of nonlinear systems with unknown time-varying parameters and control coefficients.

\end{document}